# Single-Pixel Image Reconstruction Based on Block Compressive Sensing and Deep Learning

Stephen L. H. Lau[1], Mengdi Li[1,2], Edwin K. P. Chong[3], Xin Wang[1]

[1] School of Engineering, Monash University Malaysia, Subang Jaya 47500, Malaysia.
[2] College of Optoelectronic Engineering, Changchun University of Science and Technology, Changchun, Jilin 130022, People's Republic of China.
[3] Department of Electrical and Computer Engineering, Colorado State University, Fort Collins, CO 80523, USA.

Email of corresponding author (Xin Wang): wang.xin@monash.edu

## Abstract

Single-pixel imaging (SPI) is a novel imaging technique whose working principle is based on the compressive sensing (CS) theory. In SPI, data is obtained through a series of compressive measurements and the corresponding image is reconstructed. Typically, the reconstruction algorithm such as basis pursuit relies on the sparsity assumption in images. However, recent advances in deep learning have found its uses in reconstructing CS images. Despite showing a promising result in simulations, it is often unclear how such an algorithm can be implemented in an actual SPI setup. In this paper, we demonstrate the use of deep learning on the reconstruction of SPI images in conjunction with block compressive sensing (BCS). We also proposed a novel reconstruction model based on convolutional neural networks that outperforms other competitive CS reconstruction algorithms. Besides, by incorporating BCS in our deep learning model, we were able to reconstruct images of any size above a certain smallest image size. In addition, we show that our model is capable of reconstructing images obtained from an SPI setup while being priorly trained on natural images, which can be vastly different from the SPI images. This opens up opportunity for the feasibility of pretrained deep learning models for CS reconstructions of images from various domain areas.

## 1  Introduction

In the last decade, single-pixel imaging (SPI) has emerged as an alternative to conventional digital imaging. Developed by D. Takhat *et al.* [1] at Rice University in 2006, this imaging setup provided a simple, compact, and low-cost solution that could operate efficiently across a much broader spectral range than conventional silicon-based digital cameras. SPI is an innovative imaging technique involving a simple implementation of hardware data compression at a reasonable cost compared to conventional cameras. Some works have been proposed to apply SPI on domain areas such as multidimensional imaging [2, 3], video acquisition [4, 5], and static imaging [6, 7]. The SPI architecture mainly consists of a single photodetector (SPD) and a spatial light modulator (SLM). The main working principle of SPI is that the integrated light intensity of an image imposed by a structured light pattern from an SLM is measured with an SPD.

The underlying principle of SPI is compressive sensing (CS), an effective sampling technique that requires a measurement matrix to be specified before sampling. CS demonstrates the viability of reconstructing a super-resolved signal from far fewer measurements than the Nyquist-Shannon sampling theorem [8], the tenet of signal sampling, dictates. Nyquist-Shannon sampling theorem states that the information of a signal can be preserved if uniformly sampled at a rate of at least twice its Fourier bandwidth, called the Nyquist rate. Because sampling involves measurements, the sampling rate determines the measurement cost. Unfortunately, the Nyquist rate can be too high for applications that require low measurement costs. CS was designed to capture signals that require fewer measurements than what is required based on the Nyquist sampling theorem. In CS, we measure not periodic signal samples but inner products with measurement vectors. More precisely, suppose we want to reconstruct signal $x$ with length $N$ from $M$ samples, where $M < N$. The measurement vector $y \in \mathbb{R}^{M \times 1}$ is given by $y = \phi x$, where $\phi \in \mathbb{R}^{M \times N}$ is a measurement matrix with a sampling ratio of $S = M/N$ and the rows of $\phi$ are the measurement vectors. This measurement matrix is in the form of the light projection in an SPI setup. Because the transformation from $x$ to $y$ involves a dimensional reduction, a loss in information is unavoidable. Since the dimensionality of $y$ is lower than that of $x'$ ($M < N$), there is an infinite solution for $x'$ in $\phi x' = y$, providing that no restriction is imposed on $x'$. CS provides a partial solution through the designation of the matrix $\phi$ in such a way that a sparse vector $x$ can be reconstructed approximately from $y$. To solve this ill-posed problem, CS relies on the assumption that the signal is $K$-sparse.

Many signal classes of interest exhibit sparsity or compressibility properties. For instance, smooth images are compressible in the Fourier basis, whereas piecewise images are compressible in a wavelet basis [9]. From the perspective of data acquisition, this means that the conventional method is a tremendously wasteful process, especially if we know that a signal can be expressed in $K$ basis vectors. For instance, if we take a picture with a digital camera that has millions of sensors, the image might eventually be compressed or encoded in just a few hundred kilobytes. By using the compressive sensing technique, we can ameliorate some of the drawbacks, such as the reduced overhead introduced by the encoding of the images, since we now only have to compute $K$ transform coefficients. Typically, for nonlinear reconstruction algorithms, the basis pursuit algorithm [10, 11] seeks to minimize $\|\phi x\|_1$. The reconstruction of signals can also be done via the minimization of total-variation (TV) [12]. This algorithm assumes that signals, particularly images, tend to have small total-variation compared to their energy.

Recently, deep learning algorithms showed promising results on reconstructing compressively sensed images in a simulated environment. Particularly, these deep learning algorithms are based on convolutional neural networks (CNN) for their ability to automatically extract image features to perform image-related tasks such as image classification [13, 14], object detection [15, 16], and image generation [17, 18]. CNN-based algorithms were thus extended in the CS reconstruction domain due to its superior ability in solving imaging problems. Mousavi et al. [19] proposed a stacked denoising autoencoder that models the statistical dependencies between different elements of signals and, in turn, improves signal reconstruction. Kulkarni *et al.* [20] proposed a deep learning approach that uses a convolutional neural network (CNN) that accepts CS measurements as input as outputs a reconstructed image. Due to the somewhat restrictive output image size (33 pixels × 33 pixels), they used an off-the-shelf denoiser to remove the blocky artifacts in bigger reconstructed images when multiple image patches are joined together. To solve the blocking artifact problem in the reconstructed images, Shi *et al.* [21] proposed SCSNet that learns an end-to-end mapping between CS measurements and their corresponding original images. The proposed network architecture includes a convolution layer for a learnable measurement matrix. On the other hand, Xu *et al.* [22] proposed a Laplacian pyramid reconstructive adversarial network (LAPRAN) that simultaneously generates multiple outputs with different resolutions. These methods performed well on reconstructing images from CS measurements; however, these methods are hardly applicable to an actual SPI setup. One reason is due to the unrealistic specification of the measurement matrix used in their simulated compressive sensing. For instance, most of the methods use measurement matrix consisting of real numbers. This is unfeasible for an SPI setup because each micromirror in the DMD has only binary states: ON and OFF. Therefore, the light pattern corresponding to the uploaded measurement matrix should contain either 0 or 1. Real numbers, especially negative numbers, hardly fit this important criterion. Not to mention that some methods, such as [21], involve optimizing the measurement matrix itself to achieve superior results, which makes the actual implementation even more unrealistic. Reconstruction algorithms that involve adversarial training, such as [22], can achieve a good reconstruction performance but the training can be unstable in practice.

In addition, there is a slight issue regarding the selection of dataset for training a deep learning reconstruction model. This is mainly because an SPI setup typically uses a photodiode-based sensor, instead of the silicon-based technology used in digital cameras. Besides, due to how SPI works, the data obtained is not immediately available as a proper image since it still needs to be reconstructed. Hence, obtaining training data from an experimental setup is rather unfeasible. One alternative to creating a training dataset from SPI data is by simply using a standard image dataset. However, it is unclear if these images are suitable for training. Even if it is viable, due to the plethora of image datasets, there might be a dilemma as to which dataset one should use.

In this paper, we propose a novel method for SPI reconstruction based on deep learning. The architecture of our reconstruction model is designed to take CS measurement inputs obtained from the compressive sensing of images of any dimension above the smallest default size. This allows the model to be reusable since it no longer needs to be retrained to reconstruct images of different sizes. Simulation results on multiple image datasets show that our approach achieves a better performance in image reconstruction compared to other competitive reconstruction algorithms. In addition, we have found that the model trained on datasets of natural images performed best in comparison to the models trained on simpler images, since natural images have richer semantic features. Since there is a difference between the simulated CS measurements obtained from the images used to train the model and the actual CS measurements obtained from an SPI setup, we demonstrate a method of shifting the distribution the distribution of the CS measurements to match the distribution of the training samples that the model encountered during training.

To summarize, the main contributions of this paper are as follows:
- We propose a novel CNN-based reconstruction model for CS image measurements that outperforms other competitive approaches on various image datasets.
- By training the model on a binary measurement matrix and through the shifting of data distribution, we implemented the image reconstruction of actual SPI data obtained from an experimental setup.
- The model is trained on different datasets. We made a performance comparison that shows the model performed significantly better when trained on natural images that are more complex.
- We demonstrate that our approach is capable of reconstructing images of flexible sizes as well as being robust on unseen images, opening up opportunities for pretrained CS reconstruction models.

## 2 Single-Pixel Imaging

Our SPI system is setup as follows. The light source is a monochromatic red laser (650 nm, 1 W). A digital micromirror device (DMD) that acts as a spatial light modulator is placed between the light source and the target. The beam of light is modulated by the DMD, which cast various illumination pattern on the target according to the measurement matrix uploaded by a computer. Each measurement takes around 5 seconds. The angular position of each micromirror decides the element of the measurement matrix it represents. Each micromirror has two possible states — ON and OFF —corresponding to an angle of 0 degrees and 12 degrees, respectively. Mirrors in the ON state transmit light, whereas mirrors in the OFF state reflect light away and therefore appear dark. The modulated light pattern is transmitted onto the target, which is made of a black piece of paper with cut-out sections that forms a binary transmissive object. Figure 1 shows a few examples of the targets. The lens then focuses the light (LA1740, f-85 mm) onto a single point on a photodetector (PDA36A2), which has an effective photosensitive area of ∅ 3.5 mm. A data acquisition (DAQ) device digitizes the received analog light signals before sending them to a computer for processing. The sampling rate of the DAQ (USB6001) is 20 kilosamples per second. The schematic and the real experimental setup are shown in Figure 2 and Figure 3, respectively.

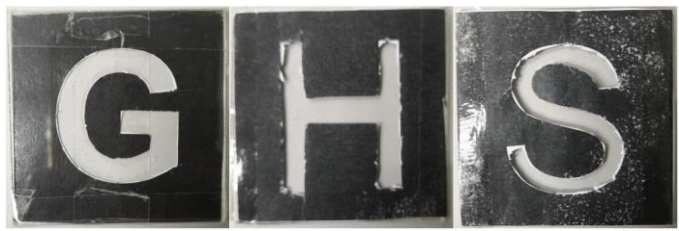

Figure 1: Targets made by sticking pattern-cut black pieces of paper on perplex. The size of each target is 6.0 cm × 6.0 cm.

The DMD we used is a Texas Instruments DMD6500, which has 1080 × 1920 micromirrors. We only activated the central area that consists of 1056 × 1056 micromirrors. For the reconstruction of the image that we obtain experimentally, we set the image size to 32 pixels × 32 pixels. Therefore, each basic unit that consists of 33 micromirrors by 33 micromirrors (1056 ÷ 32 = 33) represents each element in the measurement matrix. To reduce noise in the obtained signal, the experiment was conducted in a dark room, and the setup is enclosed by pieces of black cardboard. However, from our experiments, we found that there was still a consistent error due to the sensitive nature of the photodetector. Hence, we measured the voltage of the signal captured when it was practically completely dark and subtracted the value from our measurements.

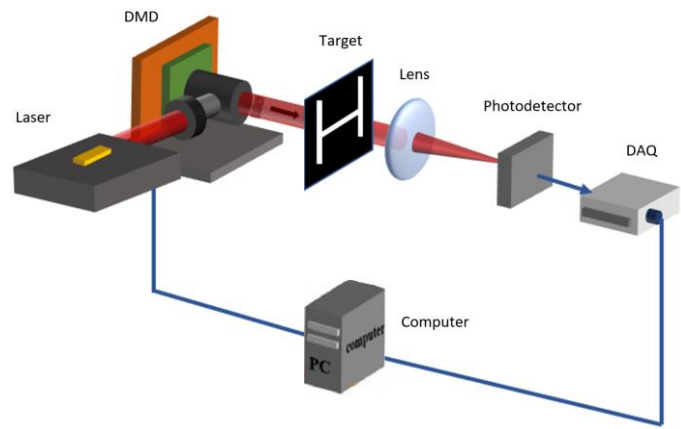

Figure 2: Schematic diagram of the single-pixel imaging setup

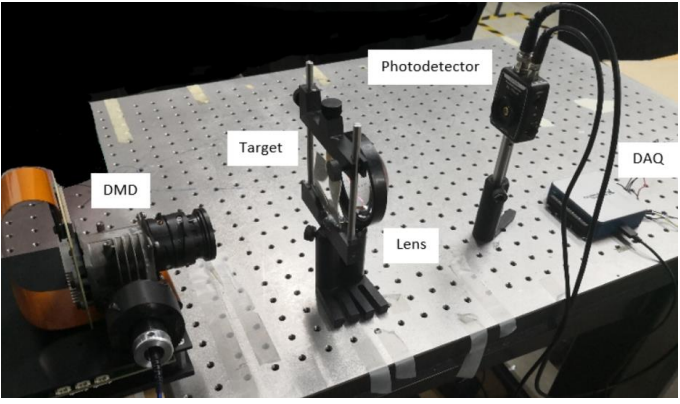

Figure 3: The experimental single-pixel imaging setup

Instead of the conventional full-image compressive sensing, block compressive sensing (BCS) [23] is used to sample measurements of the target images. The difference between the two lies in how the measurement matrices are specified, which affects how data is collected. In BCS, an image is divided into $B \times B$ blocks and sampled using a smaller measurement matrix **A** that has the same dimensionality as the block images. Suppose that $x_j$ is a vector representing, in a raster scan manner, block $j$ of input image $x$. The corresponding $y_j$ is then

$$y_j = A x_j, \tag{1}$$

where **A** is a block measurement matrix of size $M_B \times B^2$, where $M_B = \lfloor SB^2 \rfloor$. Instead of reconstructing each block image separately, we designed $\phi$ as a matrix with a block-diagonal structure:

$$\phi = \begin{bmatrix} A & 0 & \cdots & 0 \\ 0 & A & \cdots & 0 \\ \vdots & \vdots & \ddots & \vdots \\ 0 & \cdots & 0 & A \end{bmatrix}. \tag{2}$$

To make it work on a DMD, we specify matrix **A** by assigning each of its elements via a Bernoulli trial. Figure 4 shows an illustration of the light projection on the target image for both full-image CS (top row) and BCS (bottom row). In this example, we have an arbitrary target with a letter "A" as the foreground. For ease of illustration, the number of measurements in this example is 4. For the full-image CS, we require four different light projections as each corresponding measurement matrix can obtain different samples. On the other hand, BCS in this example uses only 1 block measurement matrix that covers 4 different regions on the target. Although the same block measurement matrix is used, technically speaking, they are still 4 different light projections as far as the target is concerned. Hence, one benefit of BCS is that it reduces the storage and memory requirement for the measurement matrix. We need only save the smaller measurement matrix **A** with a size of $M_B \times B^2$, rather than the full measurement matrix with a size of $M \times N$. Due to the sparsity of the matrix and the relatively low complexity of measurement matrix design, it is easy to implement in practice, such as in single-pixel imaging.

The block size we chose is $4 \times 4$. This results in a minimum sampling ratio of $\frac{1}{4 \times 4} = 6.25\%$ (since every CS measurement is responsible for $4 \times 4 = 16$ image pixels), assuming that the block covers the entire image, and the coverages are non-overlapping. The other available sampling ratios are in the multiples of the $p \times 6.25\%$, i.e., 12.5%, 18.75%, 25%, etc. We think this is a reasonable block size due to the fact that it is in the form of $2^k$ (where $k = 2$ in this case) and is therefore easily workable with any image size of $2^n$. (e.g., 64 pixels × 64 pixels, 128 pixels × 128 pixels, etc.). This choice is rather arbitrary and can potentially be changed into other block sizes. Besides, as mentioned before, the block regions are non-overlapping for simplicity purposes, but it can also be changed so that the blocks region overlap.

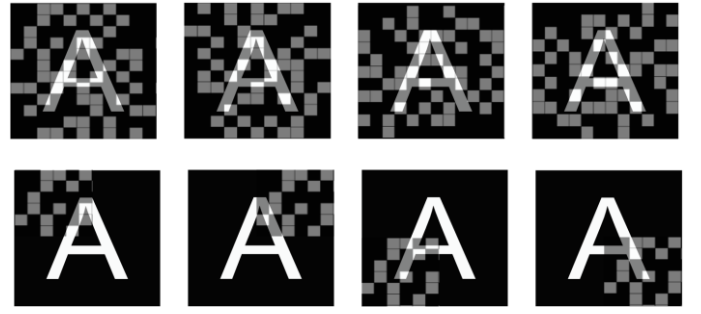

Figure 4: Difference between full-image CS (top row) and BCS (bottom row).

Figure 5 shows an example of performing BCS on an image. In general, if the image matrix has a size of $k \times k$ and the block size is $B \times B$, then the resulting CS measurements array would have $\frac{k^2}{B^2} c$ elements. The value of $c$ equals $\lfloor \alpha B^2 \rfloor$, where $\alpha$ is the sampling ratio. For the experiment, we typically first decide on the resolution of the image (in our case, it is 32 pixels × 32 pixels), then select the sampling ratio that makes $c$ a whole number. After obtaining the CS measurements, we arrange them into a 3-dimensional array of a shape of $k/B \times k/B \times c$. The first two dimensions are the spatial dimensions, whereas the third dimension refers to the new "channel" dimension.

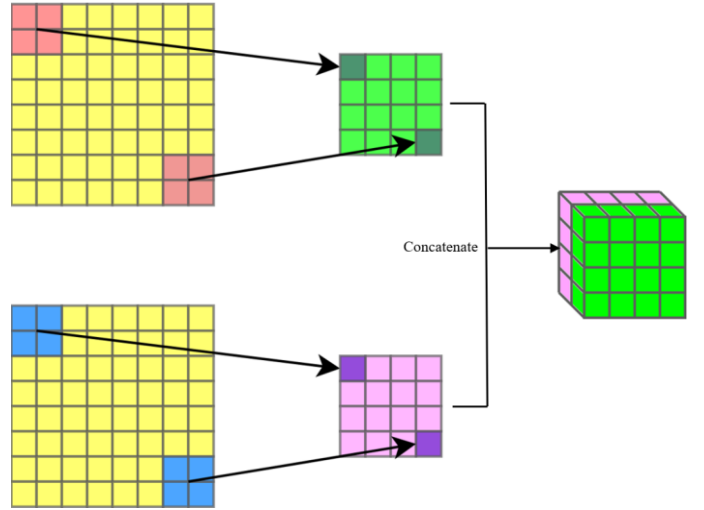

Figure 5: An example illustration of a CS measurement tensor. In this example, the size of each block is 2×2 and the size of the image (yellow grids) is 8×8. The arrows denote the dot product operation that results in a scalar number. A total of 32 measurements are to be taken. Since a maximum of 16 measurements (arranged into a 4×4 matrix each) can be obtained for each block measurement matrix (red grids and blue grids), there must be two matrices (green and purple grids) produced as a result of the block compressive sensing. The matrices are then concatenated channel-wise to form a complete CS measurement tensor, which is the input of the deep learning model.

There are two major reasons for the CS measurements to be arranged into such a structure instead of a vector. First of all, such a way of sampling an image and assigning the measurements into their relative spatial locations would allow the CS measurement array to somewhat retain the structure of the image. We wish to preserve the relationship between nearby pixels of an image, and hence we do not put the measurements into a long vector. Second, such a CS measurement array structure allows for the use of convolutional layers in the deep learning model we will propose later (see Section 3.2). Convolutional layers, as we will see later, allows for the reconstruction of images of virtually any sizes due to the nature of convolution operations.

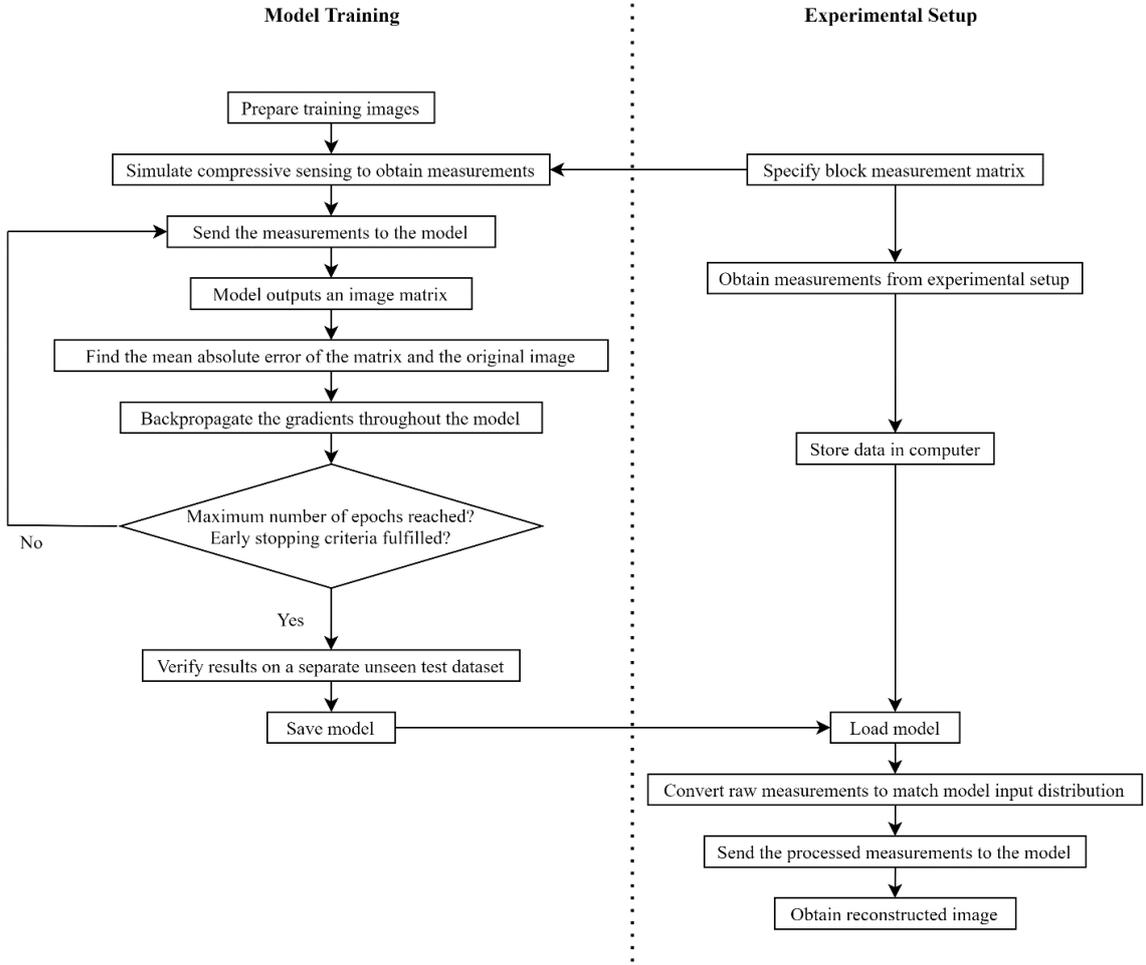

Figure 6: Overview of the full process of single-pixel imaging from data acquisition to image reconstruction.

# 3 Deep Learning Reconstruction Algorithm

## 3.1 Method Overview

As per the working principle of SPI, CS measurements are obtained through a series of structured light projection and is stored before a corresponding image is reconstructed from the measurements. In this section, we will discuss about the reconstruction method that we propose. This reconstruction method is a deep learning-based algorithm that involves a convolutional neural network (CNN). The complete workflow that combines data sampling and reconstruction is illustrated in Figure 6. In a nutshell, the measurement matrix used to obtain the CS measurements is used to simulate BCS on training images (obtained from external sources) to obtain data for model training. After the model is trained, the CS measurements obtained from the setup is post-processed and sent to the model for reconstruction.

The reconstruction model is neither rule-based nor designed via mathematical formulation. Instead, it is a parameterized function approximator that is optimized by mapping inputs (CS measurements) to their corresponding outputs (images). To obtain an optimized model, it needs to be trained on data that are obtained by simulating compressive sensing on image data. After the model is trained, inference is performed on the respective test sets of those datasets, which the model has never seen during training. This is to quantify the generalizability of the model before it can be used to reconstruct images obtained experimentally. After that, the model is deployed on the experimental setup. Each obtained measurement from the experimental setup will be converted into the correct distribution that is compatible with the model before being passed into the model. The output of the model is the image we want to reconstruct.

The rest of this section will discuss (a) the architecture of the deep learning model, (b) the datasets used to train it, (b) its training procedure, and (d) how to convert the distribution of the experimental data to match that of the training data.

## 3.2 Architecture of the Reconstruction Model

We named our proposed reconstruction model BCS-UNet. BCS-UNet is a convolutional neural network (CNN) that accepts a 3-channel CS measurement tensor[1], and outputs a matrix that corresponds to the reconstructed image. The CNN is designed to be fully convolutional; if the input has a spatial dimension (height and width) of $k/4 \times k/4$, regardless of its third dimension, then the output must be of size $k \times k$. This allows for the reconstruction of images of any size (above 32 pixels × 32 pixels, as we will soon explain). Figure 9 shows an illustration of the relationship between the sizes of the original image, the CS measurement tensor, and the reconstructed image. As we can see here, the choice of block size for BCS affects how this network is designed. In general, if the block size is $B$, then the input CS measurement tensor must have a spatial dimension of $k/B \times k/B$. For the ease of illustration, the CNN is divided into two separate CNNs, namely UpsampleNet and the UNet. In practice, these two neural networks are joined together and are jointly trained. The separation of the network is purely for descriptive purposes. Figure 10 illustrates how these two networks are combined.

---

[1] In this paper, we use the term "tensor", a term commonly used in the deep learning literature to describe multi-dimensional arrays.

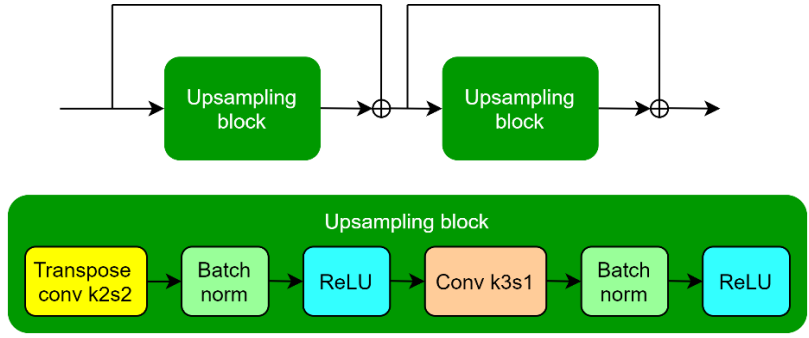

Figure 7: Network architecture of UpsampleNet. The notation $km$ and $sn$ mean the kernel size and the number of strides is $m$ and $n$, respectively.

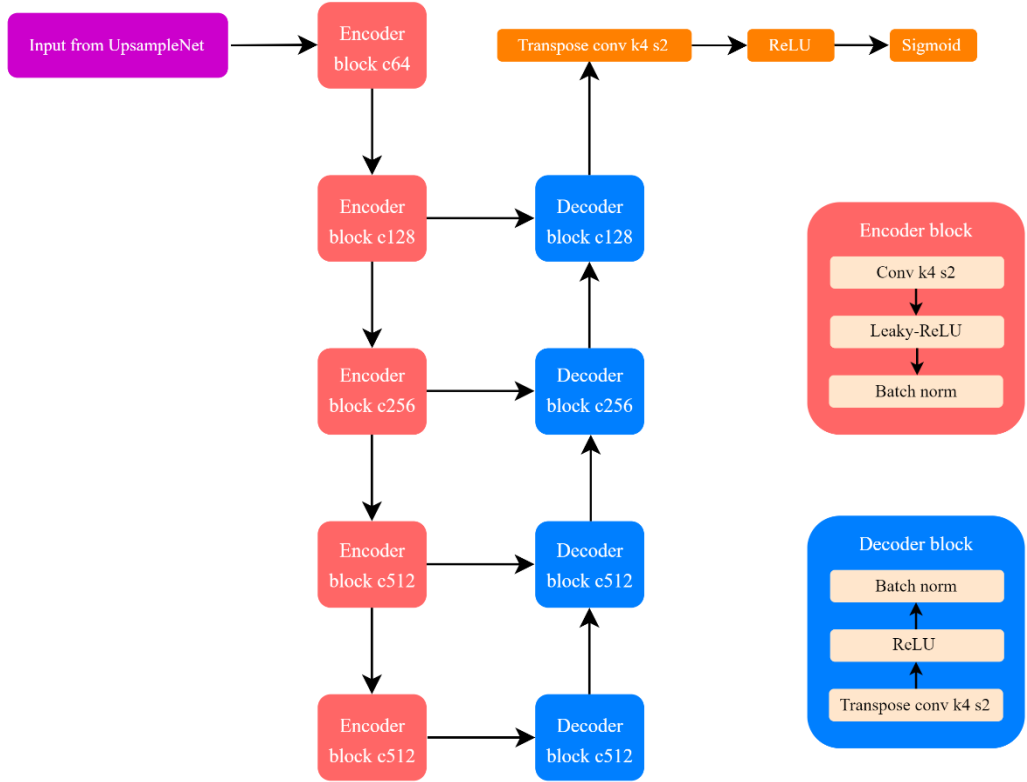

Figure 8: Network architecture of UNet. The notation of $cp$ after the encoder and decoder blocks means the number of channels of the output intermediate tensor of the block is $p$. The notation $km$ and $sn$ mean the kernel size and the number of strides is $m$ and $n$, respectively.

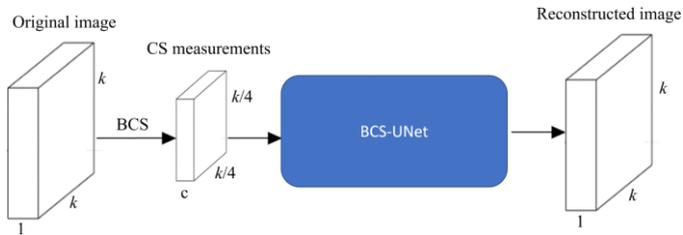

Figure 9: During the training of the reconstruction CNN, BCS is simulated on a grayscale image (size = $k \times k$) from the training dataset to obtain CS measurements that is arranged into a shape of $k/4 \times k/4$. It is then passed to BCS-UNet to output a matrix of a size of $k \times k$. The letter $c$ denotes the number of channels, which is dependent on the number of measurements (see Figure 5).

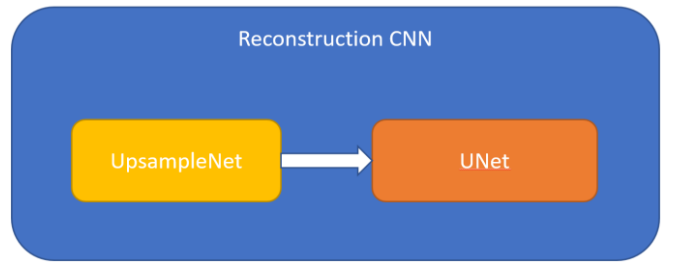

Figure 10: The reconstruction CNN consists of two sub-networks, namely UpsampleNet and UNet. The UpsampleNet takes the measurement tensor as input, and its output is sent to UNet, which in turn outputs a matrix that resembles the reconstructed image.

### 3.2.1 UpsampleNet

Figure 7 shows the architecture of UpsampleNet. UpsampleNet accepts a 3-dimensional CS measurement tensor as input. It consists of 2 upsampling blocks, each of which performs a 2× upsampling to its intermediate input. Each upsampling block consists of 1 transpose convolutional layer, 1 convolutional layer, 2 batch normalization layers [24], and 2 rectified linear unit (ReLU) layers. For each module, there is a residual connection [14] that adds the input to the output. We stack two such modules so that the spatial dimension of its output ($k \times k$) is 4 times that of the input ($k/4 \times k/4$).

### 3.2.2 UNet

Figure 8 shows a schematic diagram of UNet. The output tensor of UpsampleNet is then passed into UNet. UpsampleNet alone is insufficient in producing high-quality images. Its purpose is merely to produce a tensor that has a large enough spatial size to be the input of an encoder-decoder network such as UNet. The encoder of UNet downsamples the tensor, whereas the decoder upsamples it and eventually outputs an image we seek to reconstruct. The encoder consists of 5 stacked encoder blocks, each of which performs a 2× downsampling. Each encoder block consists of a convolutional layer, a leaky-ReLU layer (with the parameter $\alpha$ set to 0.02), and batch normalization layer. On the other hand, the decoder consists of 4 stacked decoder blocks, each of which performs a 2× upsampling. Each decoder block consists of a transpose convolutional layer, a ReLU layer, and a batch normalization layer. At the end of each encoder block, the intermediate output tensor is concatenated to the matching input tensor of the decoder block that has the same size. For instance, the output tensor of "encoder block c128" is concatenated to the input tensor of "decoder block c128", as shown in the figure. In this network, each intermediate tensor's spatial dimension decreases by a factor of 2 at every downsampling by the encoder block. Since there are 5 successive downsamplings in this network, this means that the input of the network must have a spatial dimension of at least 32 elements × 32 elements. Since the input of UNet is designed to have the same shape as its output, this implies that the size of any reconstructed image should be at least 32 pixels × 32 pixels as well. This is one of the restrictions of using an encoder-decoder network like UNet. However, we think that this image size should be a reasonable minimum, since any smaller image size contains little information anyway.

### 3.3 Datasets Used for Model Training and Evaluation

After we designed the architecture of the deep learning model, we have to train it with an image dataset. Since datasets comprising CS measurements of actual images do not yet exist, we opted to use standard image datasets as an alternative instead. Three different datasets are chosen, namely EMNIST [25], Street View House Numbers (SVHN) [26], and STL-10 [27]. The EMNIST dataset consists of images of handwritten digits and letters. The SVHN dataset consists of image patches of house numbers obtained from Google Street View images. On the other hand, the STL-10 dataset consists of natural images. The objective of using all three datasets is to determine which dataset is best suited for this project and investigate how different training datasets can affect the quality of reconstruction images. In terms of image sparsity, EMNIST looks similar to the images obtained from the binary transmissive target object. SVHN dataset contains letters that similar to the target shape in the setup as well. STL-10 images are much more complex and, therefore, have many image features from which the model can learn. Even though they are vastly different from the target objects, it should be interesting to see if the model can learn well. As mentioned previously, BCS is simulated on these images, and these resulting CS measurements are treated as model input. The images themselves act as ground truth for the model output to match. The dataset is then divided into a training dataset, a validation dataset, and a test dataset by an 8:1:1 ratio.

To virtually increase the number of training samples the model sees, various augmentations are applied to the images during model training. The augmentations include a random change in brightness and contrast, horizontal flip, and rotation at slight angles. These augmentations are turned off during the validation and testing phase, in which the model is evaluated on images it has never encountered before. In addition, all images are normalized to have a pixel range of between 0 to 1 before BCS is simulated on each image.

### 3.4 Training Procedure of the Reconstruction Model

The loss function is the mean absolute error of the output images with respect to their ground truth images, given by:

$$L = \frac{1}{K}\sum_{k}\left(\frac{1}{N}\sum_{i}|\hat{x}_i - x_i|\right) \qquad (3)$$

where $\hat{x}$ denotes the output of the model (i.e., the reconstructed image), whereas $x$ denotes the ground truth image. The symbol $|\cdot|$ denotes the modulus. The mean absolute error for each reconstructed image is summed over $N$ number of pixels in the image. The mean absolute error of the whole batch is average of all mean absolute errors across $K$ training samples in a batch, where $K$ (batch size) is set to 128. The loss function is minimised with a gradient-descent based optimiser, such as the Adam optimizer [28]. The initial learning rate is set to 0.0002, which is gradually reduced by a factor of 0.5 every time the loss does not improve.

In conjunction to model training, the model is constantly evaluated on the validation dataset that it never encounters at every training epoch. This helps with determining when the model should be stopped early before it overfits (early stopping), as well as finding the best set of hyperparameters, such as learning rate, weight decay, and various data augmentation settings. After all candidate models are trained and the best model is selected, it is then evaluated the test dataset to evaluate the performance of the model on compressive measurements extracted from unseen images in a simulated environment. The final evaluation is ideally performed only once throughout the whole project to prevent any "cheating". The purpose of having two separate datasets, validation dataset and test dataset, is to ensure that the model does not overfit to training images, and at the same time, prevent the researcher from "cheating" unknowingly through leakage of information during model training (e.g., by tweaking hyperparameter settings in an ad-hoc manner to increase the performance metrics on the unseen dataset).

### 3.5 Model Deployment on CS Measurements from Experimental Setup

After the model is trained and its performance verified on the test dataset, it is then deployed to reconstruct single-pixel images. Prior to that, some adjustments must be made for the deep learning model to work. This involves converting the input measurements obtained from the experimental setup such that they have similar distribution as that of the model training inputs. This is crucial because of how deep learning models work, and because there is a distribution difference between the training data and data obtained from the experimental setup.

Deep learning models are sensitive to the distribution of the input. It typically does well when given an input that matches the distribution of training samples, and vice versa. This is due to the fact that they are simply mapping training inputs to their respective outputs during training. Inference on the unseen test inputs only works when they have the same distribution as the training inputs. Recall that the training images are normalized to have a range of between 0 and 1 (see the last paragraph in Section 3.3). Since the model input tensors (the CS measurements) are obtained through the dot product of the training images and the block measurement matrices, the model inputs, therefore, have an inherent distribution that the CS measurements obtained experimentally (test inputs) must match.

Recall that, from the setup, each measurement is obtained by focusing all collimated all light rays, either transmitted or reflected from the target image, onto a single point where the photodetector is placed. The photodetector captures a signal as a result of the light intensity of all the light rays, but is, however, unable to record the light intensity of each spatial location. After all, that would be the whole point of single-pixel imaging. Hence, there are potentially two ways to tackle this problem: (a) ensure the voltage that corresponds to the light intensity of the light ray from each spatial location of the target to be normalized so that the resulting the distribution of the CS measurements are shifted accordingly, or (b) perform post-processing to the CS measurements to match the distribution of the training data. The former is particularly difficult to execute in practice since the CS measurements obtained are the sum of all light ray intensities (since light is focused onto the detector), and hence, the intensity of each light ray is not observable (and also defeats the idea of compressive sensing). Therefore, we chose the latter option, i.e., shifting the distribution of the experimentally obtained CS measurements. Here we show that, only with a series of linear transformations, the CS measurements obtained from the SPI setup can serve as input to the trained model during deployment, as long as they have similar distribution as that of the training data.

For the $m$-th index of the CS measurements $y$, the corresponding measurement $y_m$ can be obtained via the dot product of the $m$-th row of the measurement matrix $\boldsymbol{\phi}$ (denoted as $\boldsymbol{\phi}_m$) and the image $x$.

$$y_m = \boldsymbol{\phi}_m x$$

Although the original image $x$ is not directly observed from the single-pixel imaging setup, but we can assume that, should its elements be normalized and have a value range between 0 and 1, such a transformation can be applied to the measurements $y$ as well. Normalization is performed by subtracting the minimum of $x$ (denoted by $a$), then divide by the difference between the maximum (denoted by $b$) and the minimum.

$$x_{norm} = \frac{x - aI}{b - a}$$

where $I$ is the identity vector or matrix that has the same shape as $x$, so that the subtraction of each element of $x$ from $a$ is trivial. This normalization step introduces a shift in distribution, which can be applied to the measurements as well. Hence, we can derive the expression for $\tilde{y}_m$, the element at the $m$-th index of the transformed CS measurements:

$$\tilde{y}_m = \boldsymbol{\phi}_m x_{norm}$$
$$\tilde{y}_m = \boldsymbol{\phi}_m \left(\frac{1}{b-a}(x - aI)\right)$$
$$\tilde{y}_m = \boldsymbol{\phi}_m x \left(\frac{1}{b-a}\right) - \boldsymbol{\phi}_m I \left(\frac{a}{b-a}\right).$$

The term $\boldsymbol{\phi}_m I$ above is the dot product between the two matrices and can be simplified into the sum of all the elements of $\boldsymbol{\phi}_m$. Hence,

$$\boldsymbol{\phi}_m I = \sum_i \boldsymbol{\phi}_{m,i},$$

where the term $\sum_i \boldsymbol{\phi}_{m,i}$ denotes the sum of all the elements of the measurement matrix responsible for only measurement $m$. Since by definition $\boldsymbol{\phi}_m x$ equals to $y_m$, we finally obtain the full expression:

$$\tilde{y}_m = \left(\frac{1}{b-a}\right) y_m - \left(\frac{a}{b-a}\right) \sum_i \boldsymbol{\phi}_{m,i}$$

The expression above has four variables: the original measurement, the measurement matrix used to obtain it, the minimum and the maximum of the light intensity of the target. Hence, when an experiment is conducted, there are two extra parameters that need to be measured: the value of $a$ and $b$. The value of $a$ can be obtained by taking a measurement when all micromirrors are in the OFF state, or it can simply be set to 0 if complete darkness at the setup is assumed. The value of $b$ can be obtained when all micromirrors are in the ON state, then divided by the total number of pixels, depending on the chosen dimension of the reconstructed image. For instance, if the chosen dimension is $m$ pixels $\times$ $n$ pixels, then the value is divided by $mn$.

## 4 Simulation Results

In this section, we show some of the results obtained by reconstructing images that the model has never seen before. This is to simulate the actual scenario where we reconstruct images from any SPI data obtained from the setup. These images might be completely different from the training images, and it is, therefore, crucial for our model to be able to reconstruct them just the same. The simulation result can also help us frame the analysis into a reconstruction-only problem. That is, given that the data sampling via CS is perfect, we want to know how well a reconstruction algorithm can reconstruct SPI images.

In addition, we also benchmarked our performance against competitive reconstruction methods, namely the total-variation (TV) minimization, ReconNet, and SCSNet. The last two methods are also deep learning-based CS reconstruction methods. To ensure a fair comparison across all benchmark methods, we set up a few rules so that all methods are suitable for SPI image reconstruction, thereby ensuring the simulations reflect the realistic scenario. In particular, the measurement matrix used to simulate compressive sensing is strictly binary and unoptimizable during training. These are slight changes that do not affect TV minimization and the network architectures of ReconNet and SCSNet. However, the results reported in this paper might be different from those obtained by replicating the methods' original setup.

Table 1: Average PSNR and SSIM comparison of various CS reconstruction methods on EMNIST, SVHN, and STL-10.

| Dataset | Ratio | TV | | ReconNet | | SCSNet | | BCS-UNet | |
|---|---|---|---|---|---|---|---|---|---|
| | | PSNR | SSIM | PSNR | SSIM | PSNR | SSIM | PSNR | SSIM |
| EMNIST | 6.25 | 25.68 | 0.9511 | 26.10 | 0.9543 | 27.42 | 0.9681 | **27.43** | **0.9682** |
| | 12.50 | 27.76 | 0.9972 | 27.69 | 0.9971 | 31.34 | 0.9984 | **31.60** | **0.9986** |
| | 18.75 | 29.55 | 0.9956 | 29.67 | 0.9976 | 32.65 | 0.9993 | **32.67** | **0.9994** |
| | 25 | 31.24 | 0.9990 | 30.56 | 0.9987 | 33.86 | 0.9998 | **33.87** | **0.9999** |
| SVHN | 6.25 | 23.76 | 0.6432 | 29.44 | 0.8532 | 30.55 | 0.8909 | **31.33** | **0.9027** |
| | 12.50 | 24.31 | 0.6667 | 32.53 | 0.9122 | 36.47 | 0.9694 | **36.75** | **0.9701** |
| | 18.75 | 26.67 | 0.6945 | 32.76 | 0.9289 | 37.91 | 0.9810 | **38.94** | **0.9811** |
| | 25 | 27.54 | 0.7011 | 32.81 | 0.9411 | **39.73** | **0.9854** | 38.97 | 0.9821 |
| STL-10 | 6.25 | 22.67 | 0.6903 | 19.31 | 0.5753 | 23.37 | 0.6972 | **23.42** | **0.6981** |
| | 12.50 | 24.21 | 0.7798 | 20.34 | 0.6637 | **25.24** | **0.7852** | 25.21 | 0.7829 |
| | 18.75 | 25.21 | 0.8095 | 21.46 | 0.7134 | 26.29 | 0.8312 | **26.38** | **0.8338** |
| | 25 | 25.78 | 0.8390 | 22.37 | 0.7346 | 27.38 | 0.8632 | **27.50** | **0.8693** |
| Average | | 26.20 | 0.8139 | 28.17 | 0.8892 | 31.02 | 0.9141 | **31.17** | **0.9153** |

Table 2: Comparison of performance of various reconstruction algorithm on sample images from SVHN dataset. Sampling ratio = 25%.

| Original image | TV | ReconNet | SCSNet | BCS-UNet |
|---|---|---|---|---|
| 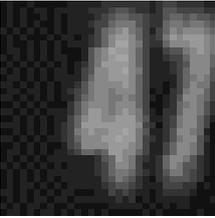 | 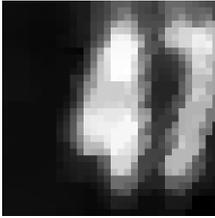 | 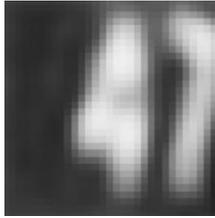 | 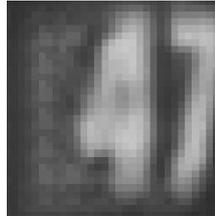 | 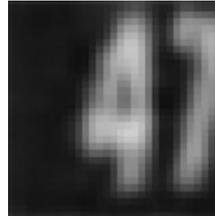 |
| PSNR / SSIM | 29.56 dB / 0.7671 | 36.74 dB / 0.9101 | 38.94 dB / 0.9811 | **38.96 dB / 0.9820** |
| 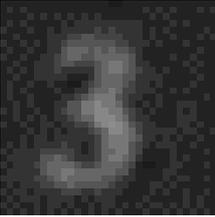 | 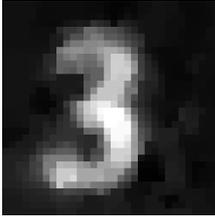 | 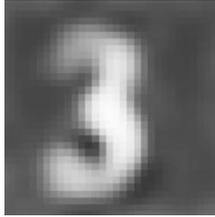 | 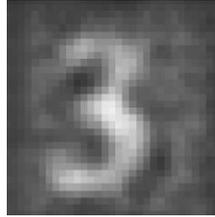 | 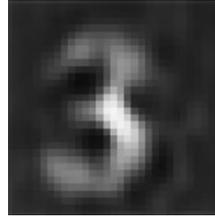 |
| PSNR / SSIM | 31.33 dB / 0.8594 | 32.64 dB / 0.9080 | 35.90 dB / 0.9509 | **37.94 dB / 0.9712** |

Table 3: Comparison of performance of various reconstruction algorithm on sample images from SVHN dataset. Sampling ratio = 12.5%.

| Original image | TV | ReconNet | SCSNet | BCS-UNet |
|---|---|---|---|---|
| 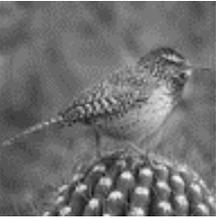 | 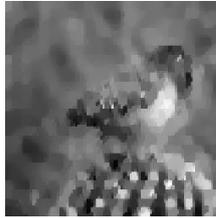 | 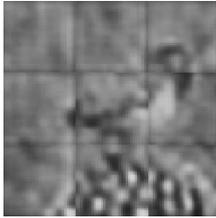 | 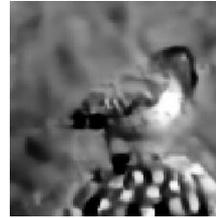 | 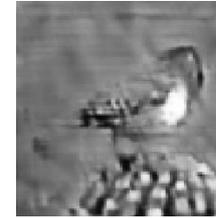 |
| PSNR / SSIM | 23.69 dB / 0.7333 | 20.72 dB / 0.6978 | 25.84 dB / 0.7806 | **25.87 dB / 0.7817** |
| 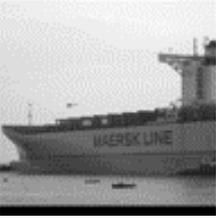 | 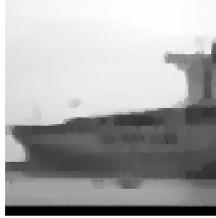 | 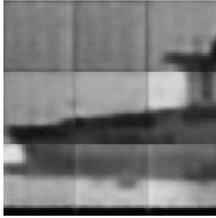 | 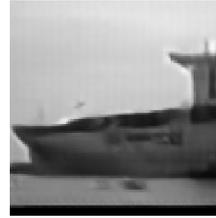 | 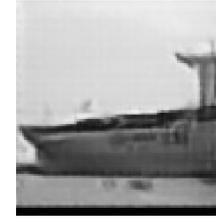 |
| PSNR / SSIM | 27.13 dB / 0.8684 | 18.64 dB / 0.5680 | 28.90 dB / 0.8809 | **28.94 dB / 0.8812** |

Table 2 and Table 3 show a qualitative comparison of reconstructed images from SVHN test dataset and STL-10 test dataset, respectively. It is apparent that BCS-UNet produced higher-quality images that do not differ too much from the original image (in terms of color and contrast) and are crisper visually. Table 1 shows the average PSNR and SSIM comparisons of different image CS reconstruction methods for various sampling ratios and datasets. As it can be seen from the table, BCS-UNet, the proposed method, outperforms a non-deep learning approach such as TV, and deep learning approaches like ReconNet and SCSNet. Although for some cases, SCSNet marginally outperforms BCS-UNet, BCS-UNet achieves a higher overall PSNR and SSIM. In a nutshell, BCS-UNet was able to generalize on unencountered images while outperforming other CS reconstruction methods.

## 5 Experimental Results

The performance of the proposed method is benchmarked against the test dataset of EMNIST, SVHN, and STL-10. Recall that, in Section 3.3, we mentioned that the model is trained separately on three different datasets, namely EMNIST, SVHN, and STL-10. The models trained on these datasets are used to reconstruct single-pixel images from the data obtained from the experimental setup. In this subsection, we are going to compare the quality of the single-pixel images reconstructed by the models trained on different datasets. For brevity purposes, the model trained on EMNIST, SVHN, and STL-10 will be referred to as "model A", "model B", and "model C", respectively.

Table 4 shows a comparison of reconstructed images using model A, model B, and model C, respectively. There seems to be a noticeable difference in reconstruction quality, though the only difference lies in the dataset on which the model is trained. For instance, although the EMNIST dataset that contains handwritten letters is more similar to the target used in the experimental setup, model A ended up reconstructing images of worse quality than model C, which is trained on STL-10, a dataset with vastly different images. One hypothesis of this observation is that the images are too sparse in nature to contain enough semantic features to be learned by the model. The reconstructed images obtained from the model B are understandably blurry because of how the dataset is curated. The images taken are originally of high-resolution but are cropped into smaller 32 x 32 patches. This blurriness is learned by the model as it attempts to reconstruct the full image while having little information due to the low sampling ratio of the measurements. Model C does better at reconstructing the images as the change in pixel values are not as jarring as the images reconstructed by model A, and the images are clearer than those reconstructed by model B. From these results, it is apparent that model C performs the best. Hence, we opted for model C to reconstruct images from the SPI data obtained experimentally.

Table 4: Comparison of reconstructed images using models trained using EMNIST, SVHN, and STL-10, respectively. Sampling ratio = 25% (number of measurements = 256).

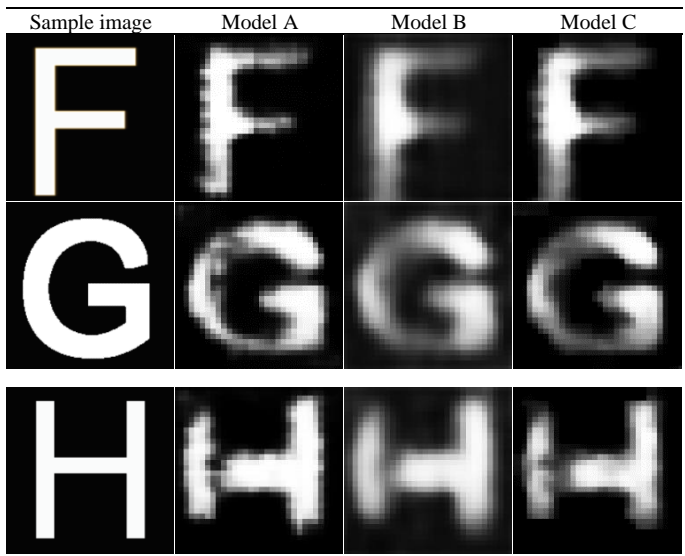

Table 5 shows a comparison of reconstructed single-pixel images across various sampling ratios, ranging from 6.25% to 25%. The difference between images is not apparent, but there is a steady improvement in terms of reconstruction quality as the sampling ratio increases. This observation is consistent with the theory that the reconstruction improves as the number of measurements increases. Table 6 shows a comparison of reconstructed images using the reconstruction methods we compared in the previous section. Similar to the simulated results, our proposed method performed significantly better than all other methods.

So far, we have demonstrated the implementation of deep learning models as a strong candidate for SPI image reconstruction. We have also showed that our model is capable of reconstructing images of a particular size that can be different from the image size of the training images. For instance, model C that we used to reconstruct those SPI images of size 32 pixels × 32 pixels were actually priorly trained on STL-10 images, which have a size of 96 pixels × 96 pixels. Note that the training step and the inference step are separate, so it is possible for the reconstructed images to have one size in training and another size in inference. The flexibility in image size is achieved through the implementation of BCS during sampling and a careful design of the architecture of BCS-UNet that is fully convolutional in nature. In addition to that, we have also demonstrated the generalizability of BCS-UNet, as it was able to reconstruct images that are different from the images used to train it. We believe our findings might open up opportunities for the reconstruction of images of other domain areas, such as magnetic resonance imaging (MRI) in the medical field. We have reported results obtained via a setup that captures binary transmissive images, but potentially it could be any type of images, as we have demonstrated the ability of our method to reconstruct various images that the model had not encountered and therefore did not overfit to (see Table 2 and Table 3). In addition, due to the ability of the model to generalize on unseen images, it is also possible to have various SPI reconstruction models that are pretrained on image datasets, since we have shown that the images can be different as long as they are rich in semantic features. The use of pretrained deep learning models has already been widely used in computer vision and natural language processing, where models are trained on large datasets and are released for practitioners to fine-tune for their separate use cases. A similar idea can potentially be implemented for SPI reconstruction models as well.

Table 5: Comparison of reconstructed single-pixel images across various sampling ratios. The reconstruction model used is Model C.

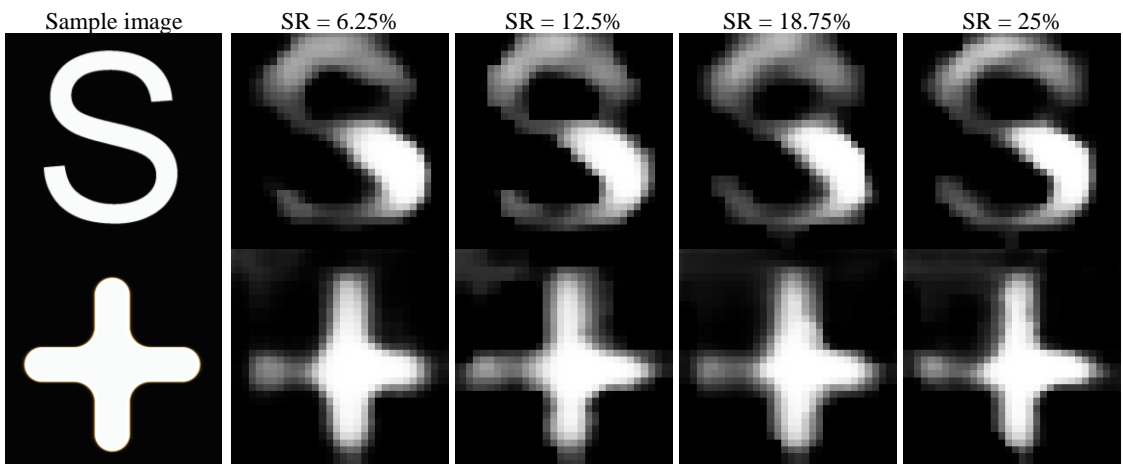

Table 6: Comparison of reconstructed single-pixel images using various methods. Sampling ratio = 6.25% (number of measurements = 64).

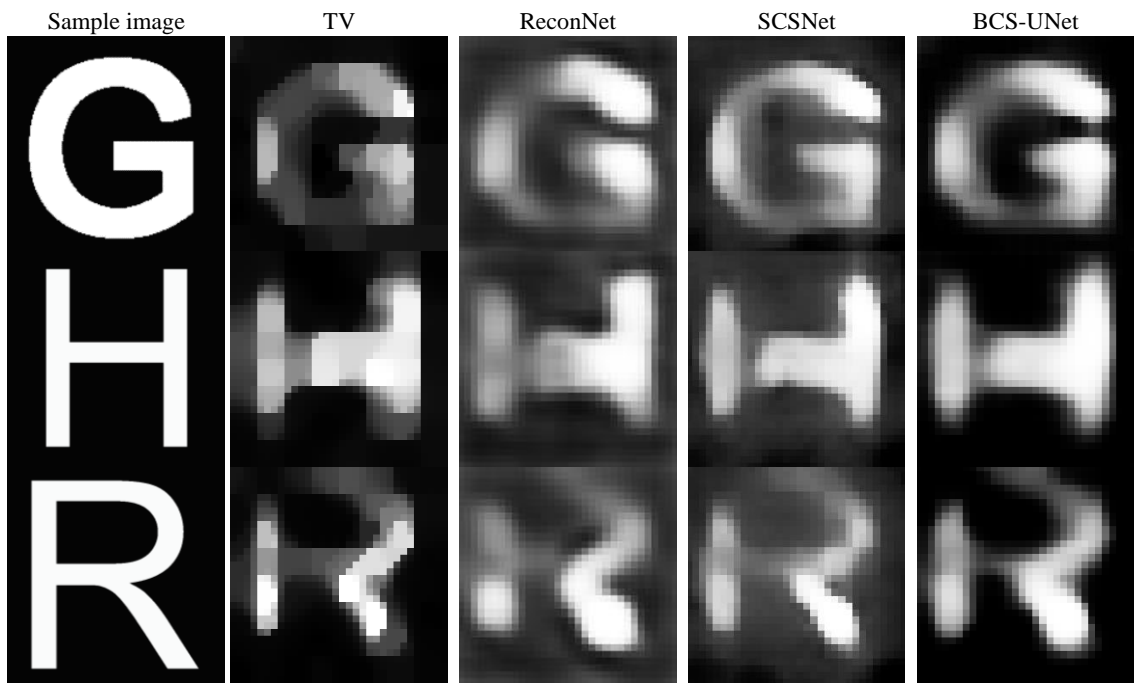

# 6 Conclusion

In this paper, we demonstrated the implementation of deep learning methods as a reconstruction algorithm for SPI images. Our approach involves the combination of block compressive sensing and a convolutional neural network-based model that outputs a reconstructed image when given a set of measurements carefully arranged in a 3-dimensional array. We also proposed a method of post-processing the SPI data to match the distribution of the training samples before sending them to the model for inference. Our approach outperforms other competitive CS reconstruction methods, both in simulation and in empirical setup. Our method is capable of generalizing on unseen images and reconstructing images of arbitrary sizes, thus making it suitable to be applied to other domain areas.